\documentclass{aa}  

\usepackage{graphicx}

\usepackage{hyperref}
\hypersetup{
    pdftitle={A Consistent 10 Myr Age for the Upper Scorpius OB Association},          
    pdfauthor={Gregory A. Feiden},   
    colorlinks=true,                                 
    linkcolor=blue,                                  
    citecolor=blue,                                  
    urlcolor=blue                                    
}

\usepackage{defs}
\usepackage{txfonts}

\titlerunning{A Consistent 10 Myr Age for Upper Scorpius}
\authorrunning{G.~A. Feiden}

\begin{document} 

   \title{Magnetic Inhibition of Convection and the Fundamental Properties of Low--Mass Stars}
   \subtitle{III. A Consistent 10 Myr Age for the Upper Scorpius OB Association}

   \author{Gregory A.~Feiden\inst{1,\,2}}

   \institute{Department of Physics \& Astronomy, Uppsala University, Box 516,
                SE-751 20, Uppsala, Sweden.
                \and
              Department of Physics, University of North Georgia, 82 College Circle, Dahlonega, GA 30597, USA. \\
                \email{gregory.a.feiden@gmail.com}
             }

   \date{Submitted: 21 October 2015; Accepted: 27 June 2016.}
 
  \abstract{
    When determining absolute ages of identifiably young stellar populations, results strongly depend on which stars are studied. Cooler (K, M) stars typically yield ages that are systematically younger than warmer (A, F, G) stars by a factor of two. I explore the possibility that these age discrepancies are the result of magnetic inhibition of convection in cool young stars by using magnetic stellar evolution isochrones to determine the age of the Upper Scorpius subgroup of the Scorpius-Centaurus OB Association. A median age of 10 Myr consistent across spectral types A through M is found, except for a subset of F-type stars that appear significantly older. Agreement is shown for ages derived from the Hertzsprung-Russell diagram and from the empirical mass-radius relationship defined by eclipsing multiple-star systems. Surface magnetic field strengths required to produce agreement are of order 2.5 kG and are predicted from \emph{a priori} estimates of equipartition values. A region in the HR diagram is identified that plausibly connects stars whose structures are weakly influenced by the presence of magnetic fields with those whose structures are strongly influenced by magnetic fields. The models suggest this region is characterized by stars with rapidly thinning outer convective envelopes where the radiative core mass is greater than 75\% of the total stellar mass. Furthermore, depletion of lithium predicted from magnetic models appears in better agreement with observed lithium equivalent widths than predictions from non-magnetic models. These results suggest that magnetic inhibition of convection plays an important role in the early evolution of low-mass stars and that it may be responsible for noted age discrepancies in young stellar populations.}
    

   \maketitle
%

\section{Introduction}
The \object{Upper Scorpius} subgroup of the larger Scorpius-Centaurus OB association is estimated to have a median age between 5 and 11 Myr. Prior to 2012, there was a broad consensus that Upper Scorpius is roughly 5 Myr old based on age determinations from stellar evolution models \citep{deGeus1992, Preibisch2002, Slesnick2008}. Crucially, \citet{Preibisch2002} showed that this age was consistent across the entire theoretical Hertzsprung-Russell (HR) diagram. However, it was quickly suggested that early-type stars in Upper Scorpius supported an older 8 -- 10 Myr age \citep{Sartori2003}. This revision was confirmed by \citet{Pecaut2012} who performed a comprehensive re-analysis of the B-, A-, F-, and G-type members of Upper Scorpius and found HR diagram ages of each stellar spectral type are consistent with an overall age of approximately $11\pm3$ Myr for the subgroup. Despite the older median age inferred from high mass members of Upper Scorpius, ages for cooler K- and M-type members remain steadfastly younger at 5 -- 7 Myr \citep[][]{Herczeg2015, Rizzuto2015, Rizzuto2015b}, incompatible with a proposed age of 11 Myr.

This problem is not unique to Upper Scorpius and is characteristic of broader issues endemic to age dating young stellar populations with stellar evolution isochrones \citep[e.g.,][]{Naylor2009, Bell2012, Herczeg2015}. HR diagram analyses by \citet{Malo2014} and \citet{Herczeg2015} show that age estimates of mid- to late-M stars in nearby young stellar moving groups are typically younger by factor of two as compared to early-M and late-K stars. On the other hand, early- to late-K stars appear to have ages that are roughly consistent with each other. However, these ages are another factor of two younger than ages typically inferred from stars with spectral type G or earlier \citep{Hillenbrand2008}. 

This picture is consistent with findings from studies of open cluster color-magnitude diagrams. Pre-main-sequence stars typically appear a factor of 2 -- 5 younger than main-sequence stars in the same population \citep{Naylor2009, Bell2012, Bell2013}. Age determinations appear to be a sensitive function of the stellar effective temperature range adopted in the analysis. The question is whether this is intrinsically real or an artifact of observational or modeling errors. 

While other young stellar populations exhibit an age gradient as a function of effective temperature, Upper Scorpius poses a particularly interesting hurdle. Previous studies of this problem are confined to the theoretical HR diagram and suffer from significant uncertainties in the accuracy of inferred stellar properties. However, observations of Upper Scorpius by \emph{Kepler}/K2 during Campaign 2 provide accurate photometric lightcurves of eclipsing multi-star systems \citep{Kraus2015, Alonso2015, David2016, Lodieu2015}. This opens up new territory for assessing hypotheses as to the origin of the noted age discordance. Models can now be constrained in the HR diagram with added constraints from the eclipsing binary mass-radius relationship. The empirical mass-radius relationship is more reliable than the HR diagram, as masses and radii can be measured directly \citep[e.g.,][]{Andersen1991,Torres2010}. Compare this to conversions from spectral types and photometric colors to \teff s and luminosities, which are subject to significant uncertainties, especially for stars with gravities intermediate between dwarf and giant scales \citep[see, e.g.,][]{Pecaut2013}. Empirical mass-radius relationships should therefore provide stringent tests of stellar model accuracy.

\citet{Kraus2015} examined stellar model agreement using the properties of a low-mass eclipsing binary, UScoCTIO5, that has two roughly 0.3 \msun\ stars with precisely measured radii. What they found was that \object{UScoCTIO5} has an estimated age of about 5 Myr if inferred from the HR diagram, consistent with the age of other low-mass stars in Upper Scorpius \citep{Preibisch2002, Slesnick2008}. Surprisingly, the age inferred for UScoCTIO5 from the mass-radius relationship implies UScoCTIO5 is about 8 Myr old, more in line with \citet{Pecaut2012}. There is some disagreement about the precise values for the masses and radii of UScoCTIO5 \citep{David2016}, but the mass-radius relationship nevertheless suggests an age of 7 -- 8 Myr. Stellar ages are not only dependent on the effective temperature of stars used in the analysis, but on the specific properties adopted. It appears that stellar evolution models are unable to provide consistent ages on multiple levels.

Unfortunately, a single eclipsing binary system with two near-equal-mass components only gives a narrow view of model reliability. More than one system is needed to draw more definitive conclusions. A second system was supplied by the characterization of the triply eclipsing hierarchical triple \object{HD 144548} \citep{Alonso2015}. HD 144548 provides three additional points that are located in a separate region of the mass-radius and HR diagrams than UScoCTIO5 (two near $1.0$ \msun\ and one near $1.4$ \msun). Characterization of HD 144548 in combination with UScoCTIO5 allows for stricter tests of model consistency, assessing not just the ability of models to reproduce individual binary stars, but the more general slope of the mass-radius relationship concurrently with HR diagram fitting.

Here, I test the hypothesis that magnetic fields may inhibit convection in low-mass pre-main-sequence stars slowing their contraction along the Hayashi track. Slower contraction times for low-mass stars would give them larger radii at a given age, pushing inferred ages to older values \citep[e.g.,][]{MM10, Malo2014}.  
In Sect.~\ref{sec:models}, I describe the model setup and method used to predict stellar surface magnetic field strengths \emph{a priori}. Section~\ref{sec:data} presents the data adopted for use in HR and mass-radius diagrams. Results about the performance of standard and magnetic stellar evolution models are given in Sect.~\ref{sec:results}, followed by  a discussion of potential uncertainties and weaknesses, as well as further evidence in support of the conclusions in Sect.~\ref{sec:disc}. Finally, the main conclusions of the paper are summarized in Section~\ref{sec:tellit}.

\section{Stellar Evolution Models}
\label{sec:models}
Models adopted in the analysis are magnetic Dartmouth stellar evolution models \citep{FC12b}. Basic physics used in these models are similar to those included in the original Dartmouth Stellar Evolution Program \citep[DSEP;][]{Dotter2008}. There are noted improvements that allow for a more accurate treatment of young star evolution \citep[see, e.g.,][]{Malo2014}. Key improvements are the explicit treatment of deuterium burning in the nuclear reaction network and the prescription of surface boundary conditions at a location in the stellar envelope where the optical depth $\tau_{5000} = 10$. 

Specification of surface boundary conditions requires the determination of the gas pressure $\pgas$ and gas temperature $T_{\rm gas}$ at a given optical depth as a function of \logg, \teff, and [$m$/H]. This is accomplished by extracting relevant quantities from stellar model atmosphere structures that are computed with the same solar abundance distribution and (hopefully) opacity data as the interior structure models. Here, model atmosphere structures are those from a custom PHOENIX AMES-COND grid \citep{Hauschildt1999a, Hauschildt1999b, Dotter2008} that adopts the solar composition of \citet{GS98} and low-temperature opacities of \citet{Ferguson2005}.

Effects related to the interaction of magnetic fields and convection, as well as magnetic fields on the plasma equation of state, are included as described by \citet{FC12b,FC13}. Radial magnetic field strength profiles are prescribed using a ``Gaussian magnetic field strength profile'' of the form
\begin{equation}
	B(r) = B(R_{\rm src}) \exp\left[-\frac{1}{2}\left(\frac{R_{\rm src} - r}{\sigma_g}\right)^2\right]
\end{equation}
where $R_{\rm src}$ is the location of the inner convection zone boundary in fractional radii and $\sigma_g$ defines the width of the Gaussian \citep{FC13}. The adopted radial magnetic field strength profile has a negligible impact on the results provided the peak magnetic energy density remains small compared to the thermal energy density \citep{FC13,FC14,FC14b}.
Peak interior magnetic field strengths are defined at $R_{\rm src} = 0.5\ R_{\star}$ for fully convective models and at the maximum of $R_{\rm src} = 0.5\ R_{\star}$ and $R_{\rm src} = R_{\rm bcz}$ (convection zone boundary) for partially convective stars. This latter condition allows magnetic models smoothly transition from a fully convective to partially convective configuration. 

The width of the Gaussian controls the magnitude of the peak magnetic field strength at the inner convection zone boundary. It was defined by \citet{FC13} as
\begin{equation}
	\sigma_g = 0.2264 - 0.1776 \left(R_{\rm src}/R_{\star}\right).
\end{equation}
\citet{FC13} provided this expression for $\sigma_g$ to keep the Gaussian localized (sharply peaked) in stars with thin convective envelopes and to make a more distributed (broadly peaked) magnetic field in fully convective stars. For stars with very thin convective envelopes or in stars where the convection envelope gradually disappears, $\sigma_g$ becomes increasingly small and produces large peak magnetic field strengths at the convection zone boundary. As a result, large magnetic pressures and magnetic energy densities are present in tenuous layers, dominating over the gas pressure in the plasma equation of state and leading to numerical convergence errors. Models with initial masses above about 1.15~\msun\ are prevented from evolving through the final approach to the zero-age-main-sequence because of this problem. To mitigate the problem, a lower limit of $\sigma_g = 0.05$ is enforced.

\begin{table}[t]
    \centering
    \caption{Equipartition magnetic field strength predictions at 10 Myr.}
    \begin{tabular*}{0.7\linewidth}{@{\extracolsep{\fill}}c c c c}
        \hline\hline\noalign{\smallskip}
        Mass    & \teff & \logg  & $\langle Bf \rangle_{\rm eq,\,surf}$ \\
        \noalign{\smallskip}
        (\msun) &   (K) & (dex)  &   (kG) \\
        \noalign{\smallskip}\hline\noalign{\smallskip}
        0.1 &   3060 &   4.16 &   2.64 \\
  		0.2 &   3261 &   4.19 &   2.51 \\
  		0.3 &   3396 &   4.20 &   2.44 \\
  		0.4 &   3517 &   4.22 &   2.41 \\
  		0.5 &   3639 &   4.24 &   2.38 \\
  		0.6 &   3760 &   4.26 &   2.34 \\
  		0.7 &   3888 &   4.28 &   2.29 \\
  		0.8 &   4031 &   4.29 &   2.22 \\
  		0.9 &   4195 &   4.30 &   2.14 \\
  		1.0 &   4397 &   4.30 &   2.04 \\
  		1.1 &   4641 &   4.30 &   1.92 \\
  		1.2 &   4910 &   4.28 &   1.83 \\
  		1.3 &   5214 &   4.25 &   1.73 \\
  		1.4 &   5569 &   4.19 &   1.58 \\
  		1.5 &   5995 &   4.06 &   1.29 \\
  		1.6 &   6618 &   3.96 &   0.94 \\
  		1.7 &   7403 &   4.15 &   0.76 \\
        \noalign{\smallskip}\hline
    \end{tabular*}
    \label{tab:equip_values}
\end{table}

The interior magnetic field strength profile requires a surface boundary condition: the average surface magnetic field strength $\langle{\rm B}f\rangle$. In previous studies, a range of values were explored and values that reproduce a given stellar fundamental property were eventually selected. 
Here, a different approach is used. Instead of adjusting models to find suitable magnetic field strengths, equipartition estimates of the maximum allowed average surface magnetic field strengths are adopted. The aim is to test the predictive abilities of magnetic stellar evolution models. Equipartition between the surface magnetic field pressure and the local gas pressure is assumed, meaning
\begin{equation}
    B_{\rm eq,\, surf} = \left(8\pi P_{{\rm gas},\, \tau = 1}\right)^{1/2}.
\end{equation}
Surface gas pressures are defined using the same model atmosphere structures used to specify surface boundary conditions. The ``surface'' is defined as the optical photosphere where $\tau_{5000} = 1$. Values of the surface equipartition magnetic field strengths are listed in Table~\ref{tab:equip_values} for stars at an age of 10 Myr. Since surface gas pressures are dependent on \logg\ and \teff, equipartition magnetic field strengths evolve toward greater values for stars undergoing quasi-hydrostatic contraction. This assumption breaks down for stars with masses above 1.55 $M_{\odot}$, whose outer convection zones disappear within about 10 Myr.

Magnetic field effects are switched on between the ages of 0.1 and 0.5 Myr years, perturbing an initial zero-magnetic-field-strength model. The precise age at which the magnetic field is switched on has no impact on the results \citep{FC12b}, provided there is a buffer between the perturbation age and ages under investigation (here, ages greater than 1 Myr). This buffer allows models to relax to a stable configuration after the perturbation. Typical peak magnetic field strengths in the models are of order 50 kG for models with $\langle Bf \rangle_{\rm eq,\,surf} \sim 2.5$~kG. Conversely, magnetic field effects are turned off during a model computation once the outer convective envelope disappears. This eventually occurs for stars with $M \gtrsim 1.15 M_{\odot}$.

\begin{table}[t]
    \centering
    \caption{Adopted solar calibration properties.}
    \begin{tabular*}{\linewidth}{@{\extracolsep{\fill}}l c c c}
        \hline\hline\noalign{\smallskip}
        Parameter & Adopted & Model & Ref \\
        \noalign{\smallskip}\hline\noalign{\smallskip}
        Age (Gyr)    & $4.567$ & $\cdots$ & 1 \\
        \msun\ (g)  & $1.9891\times 10^{33}$ & $\cdots$ & 2  \\
        \rsun\ (cm) & $6.9598\times 10^{10}$ & $\log(R/R_{\odot}) = -6\times 10^{-5}$ & 3, 1 \\
        \lsun\ (erg s$^{-1}$) & $3.8418\times 10^{33}$ & $\log(L/L_{\odot}) = -1\times 10^{-4}$ & 1 \\
        $R_{\rm bcz}/R_{\odot}$ & $0.713\pm0.001$ & $0.715$ & 4, 5 \\
        $(Z/X)_{\rm surf}$ & $0.0231$ & $0.0231$ & 6 \\
        $Y_{\odot,\, \rm surf}$ & $0.2485\pm0.0034$ & $0.2460$ & 7 \\
        \noalign{\smallskip}\hline
    \end{tabular*}
    \label{tab:solar}
    \tablebib{(1)~\citet{Bahcall2005};
    (2)~IAU 2009 \footnote{http://maia.usno.navy.mil/NSFA/NSFA\_cbe.html}
    (3)~\citet{Neckel1995}; (4)~\citet{Basu1997};~(5) \citet{Basu1998};
    (6)~\citet{GS98}; (7)~\citet{Basu2004}.}
\end{table}

Models for this study are computed using solar metallicity, [$m$/H] = 0. Values of the initial helium and heavy element mass fractions ($Y_0$ and $Z_0$, respectively) are taken from a solar-calibrated model. Similarly, a solar-calibrated mixing length parameter $\amlt$ is adopted for all models. In this way, modifications to convection are only introduced through the addition of magnetic perturbations. Solar-calibrated values are determined by finding a combination of $Y_0$, $Z_0$, and $\amlt$ that reproduce the solar radius, luminosity, and surface $\left(Z/X\right)$ at the solar age for a model of 1.0 \msun\ without a magnetic perturbation. Adopted solar quantities are given in Table~\ref{tab:solar}. Final calibration parameters are $Y_0 = 0.2755$, $Z_0 = 0.01876$, and $\amlt = 1.884$.

\section{Properties of Upper Scorpius Members}
\label{sec:data}
Empirical data compared to stellar evolution models are taken from literature sources. Properties of stars with spectral type A, F, and G are drawn from \citet{Pecaut2012}. These data were kinematically selected to be highly probable members of Upper Scorpius \citep{deZeeuw1999}. In addition, data for K and M stars are drawn from the low-mass samples presented by \citet{Preibisch1999} and \citet{Preibisch2002}. Data span logarithmic effective temperatures from about 3.5 (3\,000~K) to 4.0 (10\,000~K). From these data, median empirical stellar loci were defined using a moving median approach.

\begin{figure}[t]
    \centering
    \includegraphics[width=0.85\linewidth]{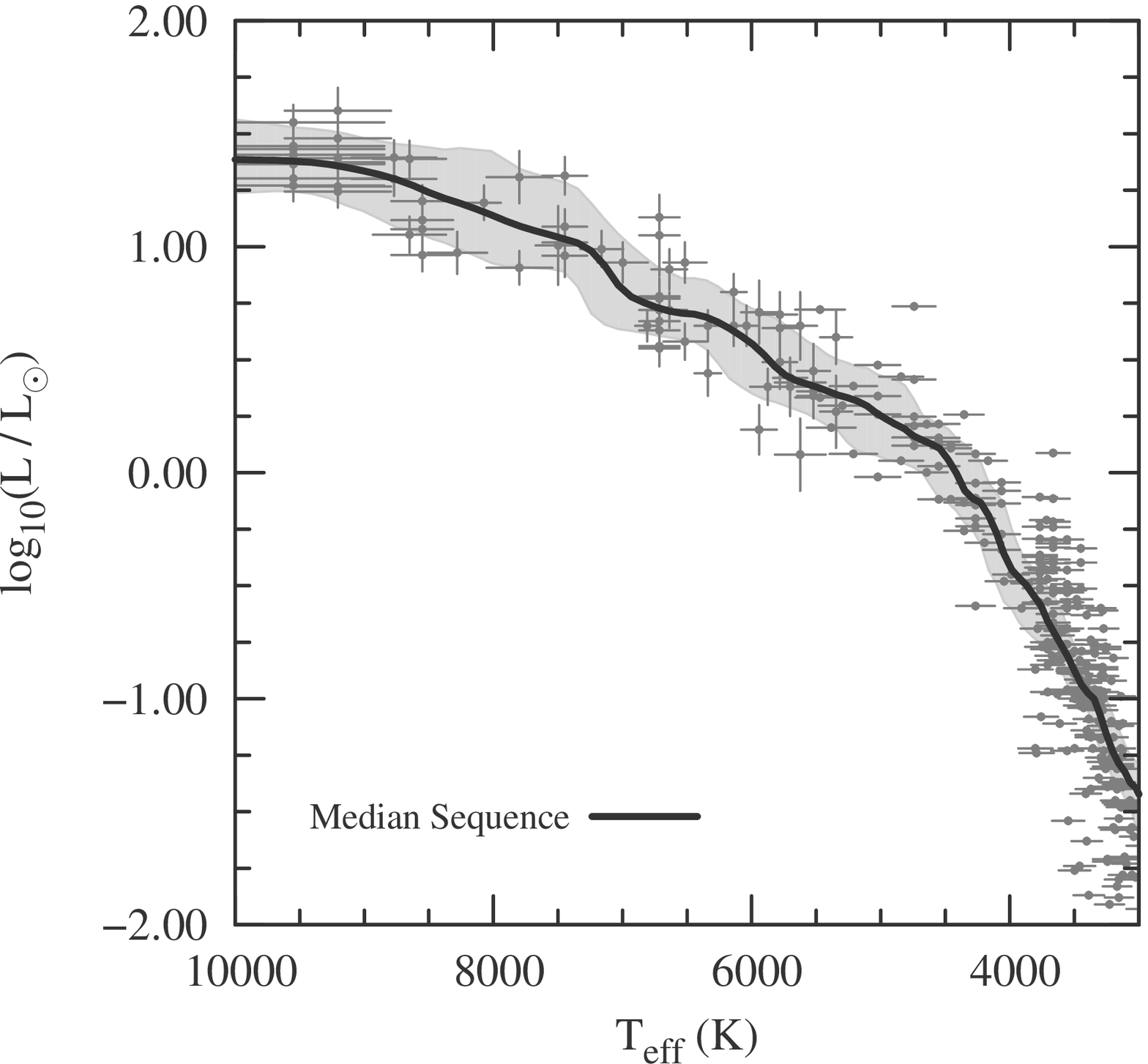}
    \caption{Hertzsprung-Russell (HR) diagram for Upper Scorpius. A-type through M-type members of Upper Scorpius \citep{Preibisch1999,Preibisch2002,Pecaut2012} are shown as gray points. A median empirical locus for the population is shown as black solid line with the 99\% confidence interval for the median value shown as a light-gray shaded region. Errorbars shown for the majority of stars with \teff~$< 5\,500$~K are prescribed as described in Sect.~\ref{sec:data}.}
    \label{fig:empirical}
\end{figure}

A moving median of logarithmic luminosities as a function of $\log$(\teff) was performed by subdividing the temperature domain into 85 bins of equal width $\Delta\log$(\teff)$ = 0.025$~dex. This width ensured at least five stars were in each bin across the full temperature domain. Bins with fewer than three stars were rejected. A median $\logLsun$\ value was estimated for each bin by finding the 50th percentile from the cumulative distribution of median values computed from a total of 10\,000 bootstrap samples. For each bootstrap sample, the $\log$(\teff) and $\logLsun$\ value for each data point was randomly perturbed by drawing values from a normal distribution centered on the measured value with a standard deviation equal to the quoted error. Thus, points were allowed to scatter in and out of neighboring bins according to their measured uncertainty. Points without uncertainties were assigned an optimistic uncertainty of 0.015~dex in both $\logLsun$\ and $\log$(\teff). Different values for the number of bins, bin widths, and number of bootstrap samples were tested and found to not significantly impact the resulting median stellar locus. The resulting median locus and corresponding 99\% confidence interval for the median value are shown in Figure~\ref{fig:empirical}.

In addition to the photometric samples, properties of two eclipsing multiple-star systems recently observed by \emph{Kepler}/K2 are included. The first of these systems, UScoCTIO5, is a low-mass eclipsing binary (\porb$= 34$ days) with a $M_A = 0.329\pm0.002$\msun\ primary and a $M_B = 0.317\pm0.002$\msun\ secondary \citep{Kraus2015}. Radii were also measured and found to be $R_A = 0.834\pm0.006$\rsun\ and $R_B = 0.810\pm0.006$\rsun\ for the primary and secondary, respectively. Extracting reliable effective temperatures for components of eclipsing binary systems is notoriously difficult. However, the similarity between the two stars in UScoCTIO5 ($F_B/F_A \approx T_B/T_A \approx 1$) allows \citet{Kraus2015} to apply several different techniques for determining component effective temperatures, including directly determining the total system bolometric flux [\fbol$= (2.02\pm0.10)\times10^{-10}\textrm{ erg s}^{-1}$]. Assuming a distance of $145\pm15$~pc \citep{deZeeuw1999}, they estimate that $T_{{\rm eff}, A, B} = 3200\pm75$~K. Properties determined by \citet{David2016} will be discussed in Sect.~\ref{sec:results}.

The second system, HD 144548 (also, HIP 78977), was recently revealed to be a triply eclipsing hierarchical triple comprised of a solar-mass close binary system orbiting an F-type tertiary \citep{Alonso2015}. The inner binary has an orbital period \porb~$= 1.63$ days, compared to \porb~$= 33.9$ days for the orbit of the inner binary with the tertiary. Through a combination of photodynamical, radial velocity, and observed minus calculated ($O-C$) modeling, \citet{Alonso2015} were able to measure masses and radii for all three stars. They find $M_A = 1.44\pm0.04$\msun, $M_{Ba} = 0.984\pm0.007$\msun, and $M_{Bb} = 0.944\pm0.017$\msun\ for the tertiary and two close binary components, respectively. Radii for the stars were measured  to be $R_A = 2.41\pm0.03$\rsun, $R_{Ba} = 1.319\pm0.010$\rsun, and $R_{Bb} = 1.330\pm0.010$\rsun. While they were unable to estimate temperatures for individual stars, they were able to estimate luminosity ratios $L_{Bb}/L_{Ba} = 0.97\pm0.05$ and $L_{Ba + Bb}/L_{A} = 0.0593\pm0.0012$. In addition, they quote an apsidal motion rate $\dot{\omega} = 0.0235 \pm 0.002\textrm{ deg day}^{-1}$, which can potentially be used as an independent age estimator \citep{FD13}.

\citet{Pecaut2012} estimated the temperature and luminosity of HD 144548 as though it were a single star. For consistency, their values of $\log$(\teff) $ = 3.788\pm0.007$ and $\logLsun = 0.83\pm0.08$ are adopted as the temperature and luminosity of HD 144548 A. A 6\% correction was applied to the luminosity estimate to remove contribution from the close binary, as established by the luminosity ratio from photodynamical modeling \citep{Alonso2015}. Complications arising with this assignment of $\log$(\teff) and $\logLsun$ are discussed in Section~\ref{sec:radius}.

Fundamental properties for the eclipsing binary EPIC 203710387 \citep{David2016, Lodieu2015} are not included in the main analysis because the component masses are uncertain at the 0.02 \msun\ level. Discussion with respect to model predictions in light the two fundamental property determinations will be presented in Sect.~\ref{sec:disc}.

\begin{figure*}[t]
    \centering
    \includegraphics[width=0.48\linewidth]{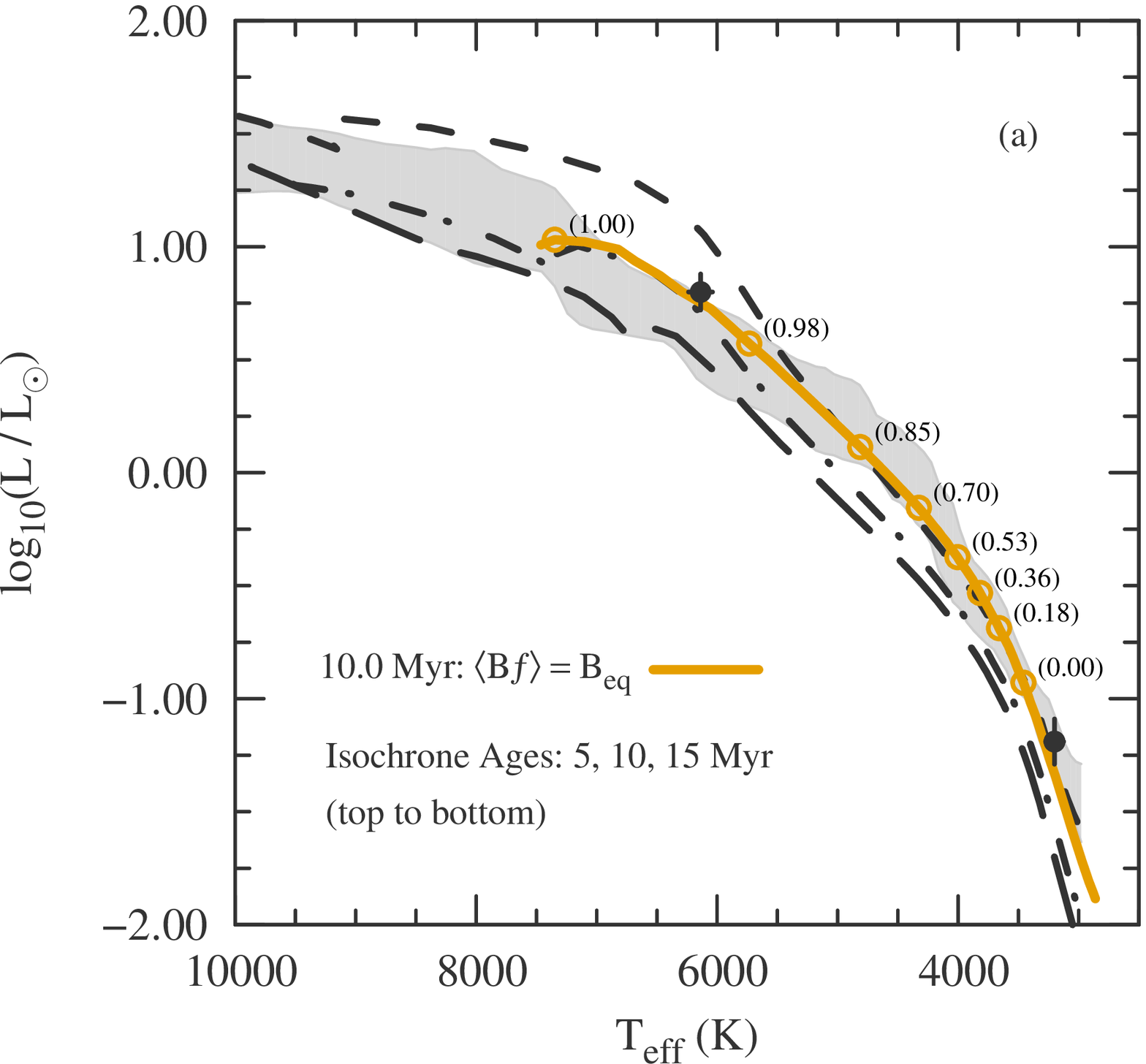} \quad
    \includegraphics[width=0.43\linewidth]{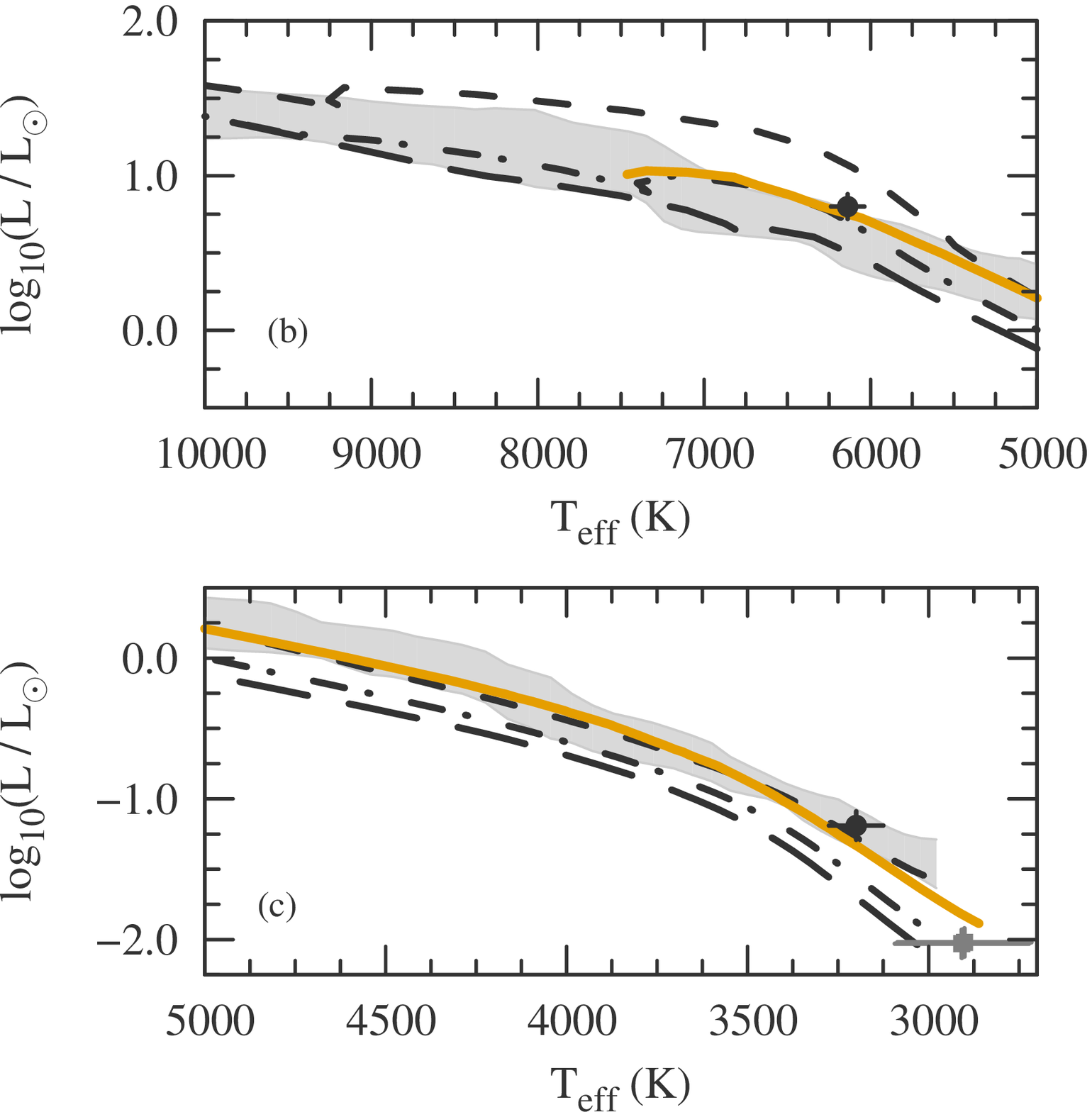}
    \caption{(a) Comparison of theoretical Hertzsprung-Russell (HR) diagram predictions to the observational locus of Upper Scorpius. The median empirical locus calculated in Section~\ref{sec:data} is shown by the light gray shaded region. Black points signify HR diagram positions of eclipsing binary members for which masses and radii are known. Over-plotted are predictions from Dartmouth stellar evolution isochrones. Standard (i.e., non-magnetic) isochrones at 5 Myr, 10 Myr, and 15 Myr are shown as short-dashed, dash-dotted, and long-dashed lines, respectively. A single magnetic isochrone at 10 Myr is plotted as a solid yellow line. Values in parentheses along the magnetic isochrone are predicted radiative core mass fractions ($M_{\rm rad.\ core}/M_{\star}$) at the circled points. All models are computed with a solar metallicity. (b) Zoom-in on the high-mass region of the HR diagram. (c) Zoom-in on the low-mass region.}
    \label{fig:ages}
\end{figure*}

\section{Results}
\label{sec:results}

\subsection{Standard Models}
An HR diagram for Upper Scorpius is shown in Fig.~\ref{fig:ages}a. Empirical data is represented by the median stellar locus (light-gray shaded region) and against three standard stellar evolution isochrones with ages of 5, 10, and 15 Myr (black dashed lines). Focusing on temperature regimes corresponding A-, F-, and G-type stars ($5\,750 <$ \teff/K $< 10\,000$), standard stellar isochrones predict ages between 5 and 15 Myr, with a median age estimate of approximately 10 Myr. 

Careful analysis of Fig.~\ref{fig:ages}b reveals that the empirical sequence closely matches models predictions at about 9 Myr where \teff~$> 8\,000$~K. Late A-type and early F-type stars with $7\,500 < $ \teff/K $ < 8\,000$ agree better with a 9 -- 10 Myr isochrone, as does the early-G-star sequence around \teff\ $\sim 5\,750$~K. In contrast, F-type stars with $6\,000 <$ \teff/K $< 7\,500$ appear marginally older and are best fit by an isochrone between 10 -- 15 Myr. These same features were noted by \citet{Pecaut2012} in their reanalysis of Upper Scorpius. Nevertheless, a 10 Myr isochrone lies almost entirely within the 99\% confidence interval for the median value across the entire range, suggesting an overall age of 10 Myr is reasonable. One can also see that HD 144548 A lies close to the 10 Myr isochrone suggesting that, at the very least, HD 144548 is approximately 10 Myr old.

However, the median age inferred from low-mass stars is about 4 -- 5 Myr. M-type stars with \teff\ $\lesssim 4\,000$~K suggest an age of 5 Myr is appropriate. The empirical stellar locus plotted in Figs.~\ref{fig:ages}a and \ref{fig:ages}c encompasses a 5 Myr standard stellar model isochrone in the M-star sequence down to \teff\ $\sim3\,100$ K. Disagreement at the lowest temperatures is likely due to a bias in the computation of the empirical stellar locus caused by a steepening of the HR diagram (see Figure~\ref{fig:empirical}).

K-type stars with $4\,000 <$ \teff/K $< 5\,250$ suggest a younger age of around 4 Myr. Due to a lack of quoted uncertainties (see Sect.~\ref{sec:data}), it is difficult to rigorously assess the significance of the disagreement between the K- and M-star ages. The position of UScoCTIO5 in the HR diagram suggests an age that is slightly younger than 5 Myr, in line with the result of \citet{Kraus2015}. Both points for UScoCTIO5 lie just above the standard stellar model isochrone at 5 Myr, but within the empirical stellar locus \citep{Preibisch1999,Preibisch2002}. Its age is thus consistent with the 4 -- 5 Myr age inferred from K- and M-type stars.

A mass-radius diagram with the two eclipsing systems is shown in Fig.~\ref{fig:mrd}. Errorbars are plotted, but are generally smaller than the individual data points. It is apparent that ages inferred for the individual binary systems are not consistent with those from an HR diagram analysis. The age of HD 144548 from the Ba/Bb components is suggested to be around 5.5 Myr according to standard stellar evolution isochrones. While an age of 5.5 Myr agrees with quoted ages for the K- and M-type stars, it is in tension with the 10 Myr age for component A from the HR diagram. Curiously, HD 144548 A lies significantly above the model mass-radius relation. Although concerning, further discussion of HD 144548 A's radius is deferred until Sect.~\ref{sec:radius}. The mass-radius diagram also suggests that UScoCTIO5 is 7.5 Myr old. This age is intermediate between the median age of the higher mass population and other low-mass stars, but it is important to note that this age is inconsistent with the HR diagram age of 5 Myr for UScoCTIO5. 

Both eclipsing systems exhibit a disagreement between their HR and mass-radius diagram age. However, the disagreements do not immediately appear to be systematic. HD 144548 appears younger in the mass-radius plane than it does in the HR diagram by roughly a factor of two. In contrast, UScoCTIO5 appears older in the mass-radius plane than it does in the HR diagram---also by about a factor of two. Slopes of the theoretical \teff-luminosity and mass-radius relationships must be altered in different directions to produce an overall agreement.

\subsection{Magnetic Models}
\begin{figure}[t]
    \centering
    \includegraphics[width=0.90\linewidth]{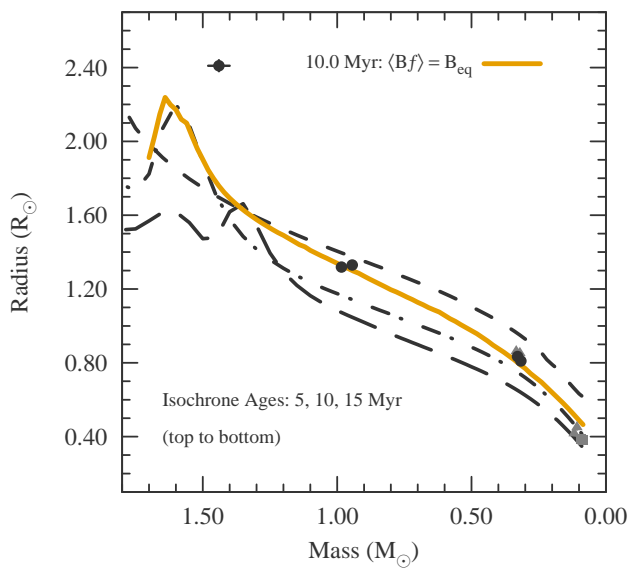}
    \caption{Mass-radius diagram for stars in eclipsing systems. Black points are from \citet{Kraus2015} and \citet{Alonso2015}, while gray points are taken from \citet{Lodieu2015} and \citet{David2016}. Model predictions for magnetic and non-magnetic isochrones are the same as in Fig.~\ref{fig:ages}.}
    \label{fig:mrd}
\end{figure}

Introducing a magnetic perturbation as described in Sect.~\ref{sec:models}, a 10 Myr isochrone is computed and overlaid in the HR and mass-radius diagrams in Figs.~\ref{fig:ages} and \ref{fig:mrd} (thick yellow solid line). An age of 10 Myr was chosen as a starting point based on the approximate median age of the A-, F-, and G-type stars determined from standard stellar evolution models. 

A 10 Myr magnetic isochrone naturally explains the factor of two age difference observed between high- and low-mass stars, provided the high-mass stars are not appreciably affected by additional non-standard physics (e.g., rotation). 
Figures~\ref{fig:ages}a and \ref{fig:ages}c (HR diagrams) show that a 10 Myr isochrone computed with equipartition magnetic fields is shifted toward cooler temperatures and higher luminosities compared to a 10 Myr non-magnetic isochrone. Inhibition of convection by magnetic fields cools the stellar surface temperature thereby slowing the contraction rate of young stars. Stars have a larger radius and a higher luminosity at a given age, as a result. The combination of cooler surface temperatures and higher luminosities makes a 10 Myr magnetic isochrone look nearly identical to a 5 Myr non-magnetic isochrone for stars with effective temperatures below about 5\,000~K. The magnetic stellar model isochrone lies on top of the 5 Myr standard model isochrone in Figs.~\ref{fig:ages}a and \ref{fig:ages}c and entirely within the empirical stellar locus for low-mass stars in Upper Scorpius. 

The magnetic model isochrone also correctly converges toward the 10 Myr standard model isochrone at warmer effective temperatures. To some degree, this  convergence reproduces an observed transition in the empirical HR diagram where the age inferred from standard model isochrones shifts from 5 Myr to 10 Myr. Below \teff~$\sim 5\,000$~K, the magnetic isochrone closely matches predictions from a 5 Myr standard model isochrone. Compare this to the small segment of the magnetic isochrone above \teff\ $\sim 6\,000$~K. Here, the magnetic isochrone traces the 10 Myr standard model isochrone. Although the model predicted transition appears to occur over approximately the correct effective temperature domain, Fig.~\ref{fig:ages}a shows a sharper transition from a magnetic to a non-magnetic sequence between $4\,500 <$ \teff/K $< 6\,000$ compared to model predictions. This is more clearly seen in Fig.~\ref{fig:empirical}, where the empirical median stellar locus exhibits a knee around \teff $\sim 4\,500$~K that roughly corresponds to the beginning of the transition region in Fig.~\ref{fig:ages}a. A distinct knee feature is absent from the magnetic model isochrone.

Moreover, inspection of the mass-radius diagram in Figure~\ref{fig:mrd} shows a 10 Myr magnetic isochrone provides reasonable agreement with the two low-mass eclipsing binary systems. It is apparent that magnetic inhibition of convection leads to more significant radius ``inflation'' at higher masses compared to at lower masses. This effect leads to a change in slope of the model predicted mass-radius relationship that is consistent with observations. Although agreement is good, it is not perfect. Depending on whether mass and radius estimates are adopted from \citet{Kraus2015} or \citet{David2016}, at 10 Myr the radii of the primary and secondary in UScoCTIO5 are either 3.2\% and 2.1\% (\citeauthor{Kraus2015}) larger than model predictions or 5.7\% and 6.6\% larger (\citeauthor{David2016}). Similarly, models at 10 Myr predict radii that are 1.4\% too large and 1.5\% too small for the Ba and Bb components of HD 144548 at 10 Myr, respectively. An age between 9.0 and 9.5 Myr is preferred for UScoCTIO5, while 10 Myr is the best fit age for HD 144548 Ba/Bb. 

This age spread could be indicative of an intrinsic age spread, but it is also possible to attribute this spread to errors in the mass and radius determinations of the eclipsing binary systems or to the fact that the low-mass stars in UScoCTIO5 may require marginally stronger magnetic field strengths than are predicted from equipartition arguments. Nevertheless, it is encouraging that models provide this level of agreement with an age and surface magnetic field strengths determined \emph{a priori}. 

\section{Discussion}
\label{sec:disc}
An age of 9 -- 10 Myr gives broad agreement across the HR and mass-radius diagrams. No stellar population in Upper Scorpius studied here robustly supports the possibility that the median age of the association is 5 Myr. There are, however, still age discrepancies that must be addressed. As noted above, the median age inferred from models for F-type stars appears to be a few million years older than the age inferred from models for stars of other spectral types. This discrepancy is perplexing and deserves further investigation, especially considering that physics potentially missing from the models would likely increase the inferred age for F-stars, widening the age discrepancy. It should be noted that most F-type stars in Upper Scorpius appear to be at a stage of evolution where models predict stellar radii increase and subsequently decrease over a period of roughly 4 -- 8 Myr. The timescale for this process is similar to the observed age discrepancy (see Fig.~\ref{fig:bump} in Sect.~\ref{sec:radius}) and may be indicative of potential shortcomings in existing model physics such as radiative opacities or nuclear reaction cross-sections. 

Age discrepancies also exist with two B-type members of Upper Scorpius: \object{$\tau$ Sco} and \object{$\omega$ Sco}. They exhibit ages between 2 and 5 Myr depending on whether rotation is accounted for in stellar evolution models \citep{Pecaut2012}. Those authors show that there are four other B-type stars that have inferred ages consistent with the 10 Myr median age proposed here, if rotation is taken into account. Notably the four stars that appear older are known binary systems while the two seemingly younger stars do not have a detected companion \citep{Pecaut2012}. The presence of two seemingly young B-type stars may be leveraged to suggest there exists a young population of stars with ages around 5 Myr thus supporting the younger age for the low-mass stellar population and thereby doing away with the need for a magnetic field explanation. However, that scenario requires that the low-mass stars form almost exclusively in a later star formation episode than the higher mass population. Although we cannot definitively rule out this scenario, we find it unlikely and maintain that the median age for the low-mass stars is too young compared to the rest of the stellar population in the absence of magnetic inhibition of convection.

\subsection{Magnetic Field Evolution \& Radiative Core Growth}
\label{sec:dynamo}
Figure~\ref{fig:ages} shows that a 10 Myr magnetic model isochrone agrees with predictions from a 5 Myr standard model isochrone for effective temperatures below about 5\,000~K, but traces a 10 Myr standard model isochrone above 6\,000~K. This suggests that magnetic inhibition of convection plays an important role in dictating the structure of cool low-mass stars, but the structure of warmer high-mass stars is relatively insensitive to the influence of magnetic fields. In between these two regimes, models predict a transition in which stars appear to exhibit a decreasing sensitivity to magnetic inhibition of convection as effective temperature increases.

As was noted in Sect.~\ref{sec:results}, Figs.~\ref{fig:empirical} and \ref{fig:ages}a indicate that a similar transition occurs over roughly the same effective temperature domain in the empirical HR diagram of Upper Scorpius. Early- to mid-G stars in Upper Scorpius (\teff$ \sim 6\,000$~K) are best described by a 10 Myr standard stellar model isochrone. Instead of continuing to follow a 10 Myr standard model isochrone, late-G stars appear to be better represented by an isochrone intermediate between 5 and 10 Myr. This causes the empirical sequence to exhibit a shallower slope than standard model predictions between $5\,750 > T_{\rm eff} / {\rm K} > 4\,500$. That the shallower slope in the empirical HR diagram and the transition predicted by magnetic stellar models are coincident, suggests that the two phenomena may be connected.

\begin{figure*}[!ht]
    \centering
    \includegraphics[width=0.42\linewidth]{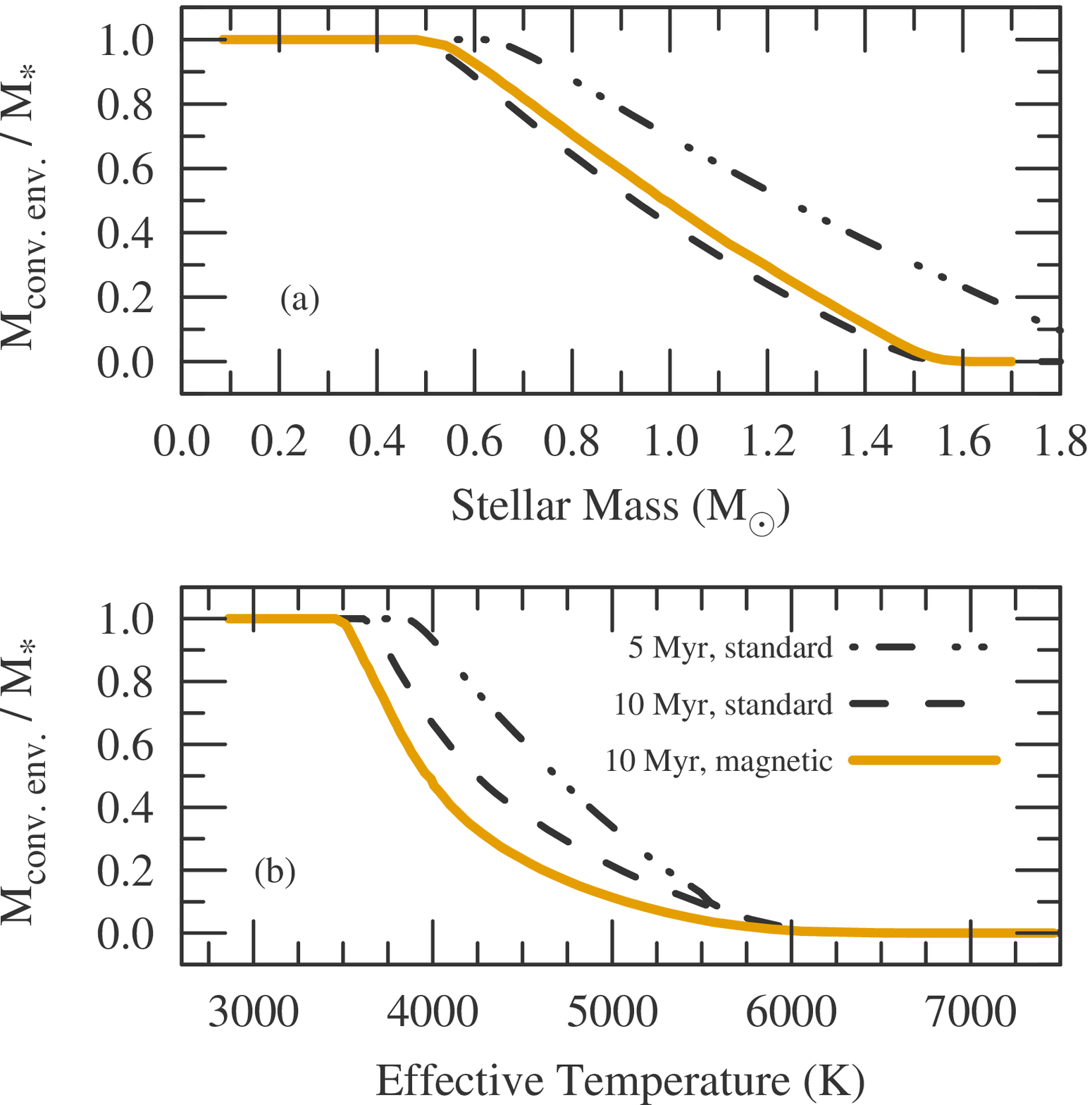} \qquad
    \includegraphics[width=0.48\linewidth]{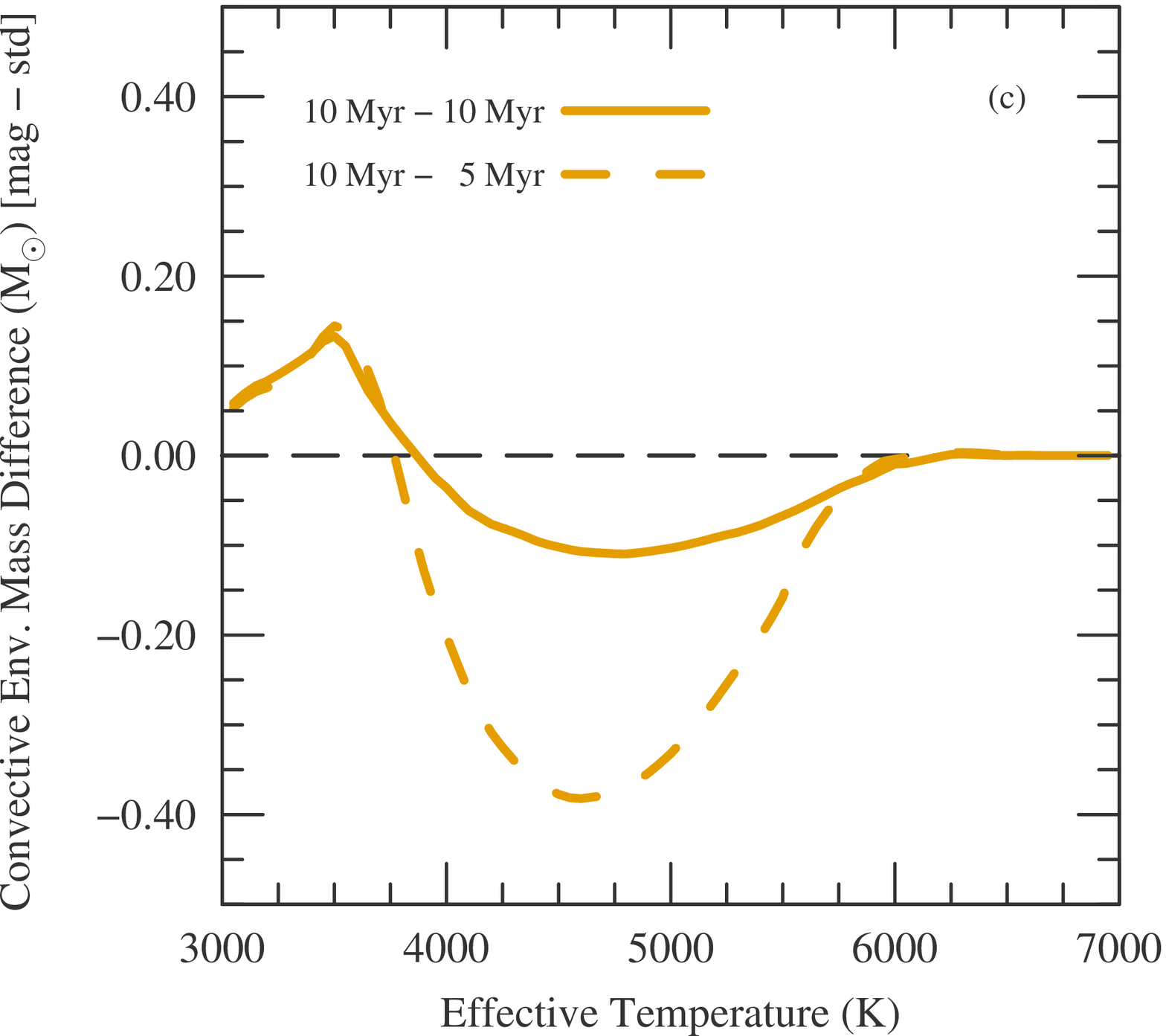}
    \caption{Mass fraction of the stellar convective envelopes as a function of (a) total stellar mass and (b) effective temperature. (c) Difference in convective envelope mass between magnetic and standard stellar evolution models as a function of effective temperature. The difference is shown as the convective envelope mass of a magnetic model minus the convective envelope mass of a standard stellar model at the same effective temperature, ($M_{\rm conv.\ env.,\, mag} - M_{\rm conv.\ env,\, std}$), shown for two ages of standard models (5 Myr, 10 Myr).}
    \label{fig:rad_core}
\end{figure*}

The transition exhibited by theoretical magnetic models can be explained by a rapid thinning of outer convective envelopes in stars with masses above about 1.2 $M_{\odot}$ \citep[e.g.,][]{Iben1965}. Once a star reaches an effective temperature of 6\,000 K, it has a very small or non-existent convective envelope, as shown in Fig.~\ref{fig:rad_core}(b). Figures~\ref{fig:rad_core}(a) and (b) show this is independent of age and mass, a fact arising as a result of the hydrogen ionization zone becoming increasingly localized near the optical photosphere with increasing effective temperature. In the absence of a convective envelope, magnetic inhibition of convection is irrelevant. Therefore, any differences observed between magnetic and standard stellar model isochrones at this effective temperature are the result of the magnetic field's influence on stellar structure at earlier evolutionary phases (i.e., delayed contraction). This provides a natural explanation for the convergence of the 10 Myr magnetic and standard model isochrones that occurs near this effective temperature in Fig.~\ref{fig:ages}(b). It then seems reasonable to posit that the shallow slope of the empirical HR diagram and the observed transition in the magnetic models trace the demise of outer convective envelopes and the growth of radiative cores in young, pre-main-sequence stars.

While the demise of outer convection zones explains the final convergence of magnetic and standard models, it doesn't necessarily explain the beginning of the transition that occurs between $T_{\rm eff} = $ 4\,500~K and 5\,000~K.  In this transition region, magnetic models have masses between 1.26 $M_{\odot}$ ($T_{\rm eff} = $ 4\,500~K) and 1.40 $M_{\odot}$ ($T_{\rm eff} = $5\,000~K). Magnetic models predict that these stars have convective envelope masses between 24\% and 12\% of their total stellar mass, respectively (see Figure~\ref{fig:rad_core}a). The 1.40 $M_{\odot}$ model is of particular note, as Figure~\ref{fig:mrd} suggests that 1.40 $M_{\odot}$ stars are entering a brief period of radius inflation that accompanies the increasing energy generation from the $p$-$p$ chain and CN-cycle burning prior to the onset of core convection \citep[e.g.,][]{Iben1965,Bodenheimer1965}. Photospheric gas pressures rapidly drop during this phase of evolution bringing about a rapid drop in surface equipartition magnetic field strengths. As expected, a rapid decrease in equipartition magnetic field strengths is observed among the data in Table~\ref{tab:equip_values} beginning around 1.40 $M_{\odot}$. Stars above 1.40 $M_{\odot}$ become less sensitive to magnetic inhibition of convection because their magnetic fields are intrinsically weaker.

If the transition region begins around 4\,500~K, as one might expect from the knee in the empirical HR diagram (Fig.~\ref{fig:empirical}), a rapid decrease in surface magnetic field strengths cannot fully explain the start of the transition. Note that Fig.~\ref{fig:rad_core}(c) suggests that the mass difference between convective envelopes in magnetic and standard stellar models occurs around $T_{\rm eff} \sim 4\,500$~K,\footnote{The slope in the convective envelope mass difference below $3\,500$~K is due to the fact that models predict stars are fully convective below that temperature, but cooling of the stellar photosphere by magnetic inhibition of convection means magnetic models have a larger mass at a given $T_{\rm eff}$ than standard models.} suggesting there is something relatively unique occurring at that effective temperature. However, Fig.~\ref{fig:ages} shows no noticeable change in morphology at  $T_{\rm eff} \sim 4\,500$~K. Another mechanism must be responsible for the failure of magnetic models to reproduce the sharp transition.

The sharp knee at \teff~=~4\,500~K in Fig.~\ref{fig:empirical} may be the result of a global change in magnetic field topology. Evidence for this comes from \citet{Gregory2012}, who noted that large-scale magnetic field topologies observed on pre-main-sequence stars appear to correlate with their predicted \emph{radiative core mass}, or alternatively, their convective envelope mass. Specifically, \citet{Gregory2012} noted a shift from predominantly dipolar axisymmetric magnetic field topologies to multi-polar non-axisymmetric field topologies when the estimated radiative core mass $M_{\rm rad.\ core} > 0.4\ M_\star$. It was noted above that at \teff~=~4\,500~K, magnetic models predict a star with $M_{\star} = 1.26\ M_{\odot}$ and $M_{\rm conv.\ env.} \approx 0.24\ M_{\star}$ ($M_{\rm rad.\ core} \approx 0.76\ M_{\star}$), well below the predicted convective envelope mass where stars are expected to shift toward multi-polar non-axisymmetric field topologies.

However, \citet{Gregory2012} used non-magnetic models \citep{Siess2000} to establish a relationship between magnetic field topology and stellar interior structure. Hypothetically, if they were to observe a star in Upper Scorpius with \teff~=~4\,500~K, their HR diagram analysis would suggest that the star is approximately 5 Myr old. Figures~\ref{fig:rad_core}(a) and (b) reveal that a star with \teff~=~4\,500~K at 5 Myr is expected to have a mass of $M_{\star} = 1.10\ M_{\odot}$ with $M_{\rm conv.\ env.} \approx 0.67\ M_{\odot}$ and thus $M_{\rm rad.\ core} \approx 0.43\ M_{\odot}$ (see Fig.~\ref{fig:rad_core}(a)). This radiative core mass corresponds to 39\% of the total stellar mass and is consistent with the boundary identified by \citet{Gregory2012}.

The implications of this results are that: (1) the knee observed in Fig.~\ref{fig:empirical} may be the result of a change in magnetic field topology and (2) the shift in magnetic field topology from a predominantly axisymmetric configuration to a non-axisymmetric configuration may occur when the radiative core mass is much larger---and convective envelope mass much smaller---than initially suggested. Magnetic models suggest the shift in global magnetic field topology may occur when $M_{\rm conv.\ env} / M_{\star} \approx 0.25$ ($M_{\rm rad.\ core} / M_{\star} \approx 0.75$) as compared to $M_{\rm conv.\ env.} / M_{\star} \approx 0.60$ \citep[$M_{\rm rad.\ core} / M_{\star} \approx 0.40$;][]{Gregory2012}.

This discussion is premised on the reality of the apparent transition at \teff~=~4\,500~K and the demise of convective envelopes producing the shallower slope in the empirical HR diagram. The shallower slope may be an artifact of how the empirical locus was computed. The empirical HR diagram is somewhat under-populated in the immediate vicinity of \teff~$= 5\,500$~K, raising the possibility that the running median has artificially created a smooth transition between warmer and cooler stars. However, the HR diagram is well-populated at slightly warmer and slightly cooler temperatures, suggesting the two regions must be somehow connected. While the precise morphology of the HR diagram in this region may be subject to alteration with additional data, the general trend should be robust against errors in how the empirical locus is computed.

Errors transforming from photometric colors or spectral types to effective temperatures may provide a plausible explanation for the shallower slope of the HR diagram in the vicinity of 5\,000~K. Below this temperature, stellar atmospheric opacity becomes increasingly dominated by molecular species, particularly TiO in the optical and H$_2$O in the infrared, which poses problems when attempting to derive a robust effective temperature scale. This is particularly problematic for the HR diagram shown in Fig.~\ref{fig:empirical} because stars above \teff$ \sim 5\,000$~K were transformed from the observational to theoretical plane by different authors (\citealt{Pecaut2012} at warmer temperatures and \citealt{Preibisch1999} at cooler temperatures). If the effective temperature scale adopted by \citet{Preibisch1999} for cool stars is too cool by about 200 K, the observed transition might disappear and low-mass K- and M-type stars would fall closer to the 10 Myr standard model isochrone.

\citet{Preibisch1999} constructed an spectral type to effective temperature transformation at a gravity intermediate between dwarf and giant scales using the transformation derived for luminosity class IV objects by \citet{deJager1987}. This temperature scale is consistent with the more recent scale of \citet{Herczeg2015} who used PHOENIX BT-Settl model atmospheres \citep{Allard2011} to infer effective temperatures from photometric colors. Agreement between empirical methods and theoretical predictions is reassuring. However, comparing these results to the 5 -- 30 Myr pre-main-sequence star temperature scale proposed by \citet{Pecaut2013}, one finds that the temperature scale adopted by \citeauthor{Preibisch1999} is about 100~K cooler for early K-type stars, but the two temperature scales agree around spectral type M0. For early-to-late M-type stars, \citeauthor{Preibisch1999} find temperatures about 100~K warmer than \citet{Pecaut2013}, a reversal from the comparison for late-K-type stars. Considering these differences, the sharp knee feature in the HR diagram may be accentuated by the adoption of an erroneous temperature scale, but it does not appear sufficient to fully explain the shallower slope observed in the HR diagram.

The observed knee and shallower slope in Fig.~\ref{fig:empirical} appear to be genuine features. General agreement between the behavior of the magnetic model isochrone and the observed stellar locus provides evidence that magnetic inhibition of convection is the mechanism that is producing these effects. This means that magnetic fields may be responsible for the age discrepancy between the high- and low-mass stellar populations in young associations. However, it is clear that this statement only holds with respect to the plausible identification of a general mechanism as the models fail to reproduce the precise morphology at the ``magnetic sequence turn-off.'' This is perhaps because magnetic models do not account for shifts in global magnetic field topology. If the age difference between warmer and cooler stars in Upper Scorpius is the result of magnetic inhibition of convection, the transition region between $4\,500 <$ \teff/K $< 6\,000$ provides an excellent laboratory for studying magnetic dynamo evolution as a function of stellar surface convection zone properties.

\subsection{Initial Mass Function}
The presence of magnetic and non-magnetic sequences has implications for the (sub)stellar initial mass function of Upper Scorpius \citep{Ardila2000, Preibisch2002}. Section~\ref{sec:results} described how magnetic inhibition of convection predominantly shifts stars of a given mass to cooler temperatures. As a result, stars with a given effective temperature and luminosity have a higher mass when magnetic inhibition of convection is considered. Furthermore, the presence of a transition region characterized by gradually declining influence of stellar magnetic fields that links the magnetic and non-magnetic sequences will tend spread stars with similar masses across a larger effective temperature domain. 

For example, consider the predicted mass difference between a star with $T_{\rm eff} = $ 6\,000~K and one with $T_{\rm eff} = $ 4\,500~K. Non-magnetic models predict a mass difference of 0.85 \msun\ and 0.50 \msun\ at 5 Myr and 10 Myr, respectively. However, magnetic models suggest that the mass difference is 0.25 \msun\ at 10 Myr. This is at least a factor of two smaller than standard model predictions.

Neglecting magnetic inhibition of convection would produce stellar mass distributions that show a paucity of stars at higher masses compared to standard predictions for field and cluster initial mass function \citep[e.g.,][]{Salpeter1955, Kroupa2002, Chabrier2003}. \citet{Preibisch2002} note a possible excess of low-mass stars compared to field initial mass functions \citep{Scalo1998, Kroupa2002}, but it's not immediately clear whether this potential excess of low-mass stars is fully consistent with predictions from magnetic model predictions. Assessing the impact of magnetic inhibition of convection on the (sub)stellar initial mass function of Upper Scorpius is reserved for future work.

\subsection{Radius of HD 144548 A}
\label{sec:radius}
\begin{figure}
    \centering
    \includegraphics[width=0.85\linewidth]{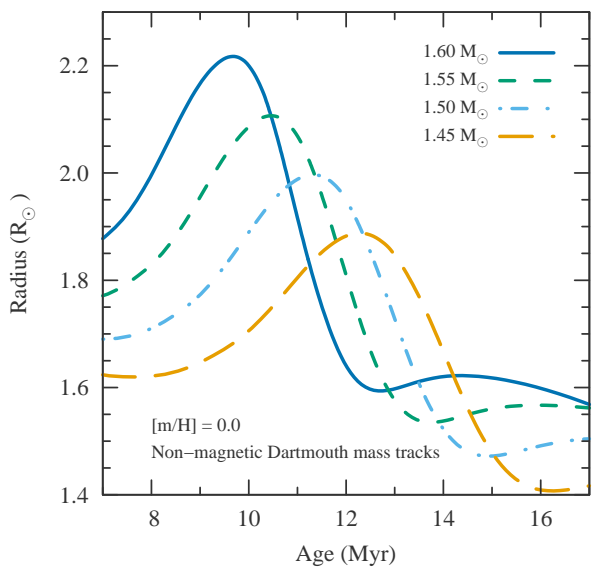}
    \caption{Brief period of radius inflation described in Sect.~\ref{sec:radius} for stars with masses of 1.60\msun\ (blue solid), 1.55\msun\ (green short-dashed), 1.50\msun\ (blue dash-dotted), 1.45\msun\ (yellow long-dashed). Mass tracks are computed without a magnetic field (i.e., non-magnetic configuration). Effective temperatures at these masses are characteristic of F stars at 10 Myr.}
    \label{fig:bump}
\end{figure}

Figure~\ref{fig:mrd} reveals that the radius of HD 144548 A is significantly larger than is expected from standard stellar evolutionary predictions. This poses a problem to claims of consistent ages across stellar populations in Upper Scorpius. However, it is not immediately clear that the observed radius error is a true astrophysical problem. For instance, adopting the effective temperature and luminosity estimate from \citet{Pecaut2012}, one derives a radius of $R = 2.2 \pm 0.2 R_{\odot}$. This is formally consistent with the radius derived by \citep{Alonso2015}, but the mean value is more consistent with estimates from stellar evolution isochrones around 10 Myr. Stellar models show an increase in stellar radius for stars with masses in the vicinity of $M = 1.5$\msun, illustrated as a function of stellar mass and age in Fig.\ \ref{fig:bump}. As described in Sect.\ \ref{sec:dynamo}, this radius increase is due to the convective envelope responding to increased energy generation from CN cycle ignition.

Assessing in detail whether the larger radius of HD 144548 A is real is beyond the scope of this work. However, if it requires revision, the result is unlikely to affect the inferred masses and radii derived for the close binary. Those largely depend on photodynamical modeling of the \emph{Kepler}/K2 lightcurve during events where Ba and Bb eclipse one another, not where the pair eclipse the tertiary component. Results in Sect.~\ref{sec:results} should be robust. Still, confirmation of the star's radius would be beneficial for future investigations to ensure there are no adverse effects on the radii of the B components. 

HD 144548 appears to be in a fairly rapid phase of evolution given either the radius estimate from \cite{Pecaut2012} or \citet{Alonso2015}. This phase of evolution, described in Section~\ref{sec:dynamo}, is sensitive to radiative opacities, individual abundances of carbon and nitrogen, and nuclear reaction cross sections. Radiative opacities and $p$-$p$ chain reaction cross sections determine when radius inflation begins, while CN-cycle reactions largely control when radius inflation ends. The maximum radius reached by a star during this phase of evolution is sensitive to the adopted CN-cycle cross sections and individual abundances of carbon and nitrogen. Therefore, it may be possible to use HD 144548 A to constrain these physics or to constrain the C/N ratio in Upper Scorpius. This requires that the properties of HD 144548 A, and several other F-type stars, be accurately determined.

Perhaps there are other physics missing for HD 144548 A that result in the radius discrepancy. Magnetic inhibition of convection does not appear to play a significant role in governing the radius at this phase of evolution. Convective envelopes contain very little mass ($M_{cz} < 10^{-3} M_{\odot}$) and models indicate that, during the brief period radius inflation, magnetic and non-magnetic mass tracks tend to converge (see Fig.~\ref{fig:mrd}). However, the exact role that magnetic fields may play in governing the pre-main-sequence contraction of young A-, F-, and G-type stars remains relatively unexplored. While magnetic fields are included in models of these stars, the equipartition magnetic field strength is fixed to the predicted value at 10 Myr. Stronger surface magnetic fields may exist at younger ages, which might delay contraction and lead to larger radii at a given age. Initial tests suggest this is likely not a significant factor in the observed radius discrepancy between models and HD 144548 A.

Effects of strong magnetic fields on convection in stellar cores are also neglected. Strong magnetic fields may be generated by the interaction of a fossil field and vigorous dynamo action \citep[e.g.,][]{Featherstone2009}; a scenario supported by the observation of suppressed asteroseismic dipolar and quadrupolar modes in the cores of a fraction of evolved intermediate mass stars \citep{Fuller2015,Stello2016a,Cantiello2016}. Core magnetic fields in stellar evolution models used in this work are below 1 G, which is insufficient to influence core convection. The impact of significantly stronger magnetic fields on core convection---especially the onset of core convection---and the subsequent effect on the properties of young intermediate mass stars is reserved for a future investigation. 

\subsection{On Starspots}
Starspots have been suggested as an alternative mechanism to slow the contraction of young low-mass stars \citep{Jackson2009, MM10, Jackson2014a, Somers2015b}. Assuming starspots are analogous to sunspots, spots can be understood to be local manifestations of convective inhibition near the optical photosphere \citep{Biermann1941,Deinzer1965}. In other words, the two phenomena are closely related. However, the primary difference between theoretical treatments starspots and a more general magnetic inhibition of convection in stellar evolution models is that spots are assumed to be strong localizations covering some fraction of the stellar surface whereas magnetic inhibition of convection is assumed (for simplicity) to be globally pervasive. Starspots have qualitatively and quantitatively similar effects on stellar fundamental properties \citep[radius, \teff, luminosity;][]{Spruit1982a,Spruit1986,Somers2015b}, but differ in their impact on photometric properties. Colors of spotted stars are the result of a superposition of surface regions of different effective temperatures \citep{Spruit1986}, whereas a global inhibition of convection still assumes a single temperature optical photosphere \citep{Jackson2014a}.

Agreement between ages and stellar fundamental properties of low-mass stars predicted by magnetic stellar evolution isochrones presented in this work with ages of higher mass stars in Upper Scorpius may indicate that a global inhibition of convection accurately captures relevant physics involved in halting contraction of low-mass stars. If starspots are also to be considered, one might expect to find that young stars in Upper Scorpius have a high coverage fraction of spots ($f \sim 1$). In this limit, a global inhibition of convection and spotted models \citep[e.g.,][]{Somers2015b} should provide equivalent explanations for observed HR and mass-radius diagram discrepancies.

Photometric brightness modulation is observed in lightcurves of UScoCTIO5 and HD 144548, suggesting spots are present on their optical photospheres. However, it may be that a large fraction of the surface is covered by a non-zero magnetic field of near-equipartition value, accompanied by smaller regions of strong local magnetic flux concentration, such as observed on the young star \object{AD Leo} \citep{Shulyak2014}. The latter can lead to the appearance of cooler spots, although only covering a small fraction of the stellar surface. If spots have a small areal coverage, it is more likely that they will have a negligible impact on stellar structure as compared to global inhibition of convection from the more ubiquitous non-zero background magnetic field.

\subsection{Lithium Depletion}

\begin{figure}[t]
    \includegraphics[width=0.85\linewidth]{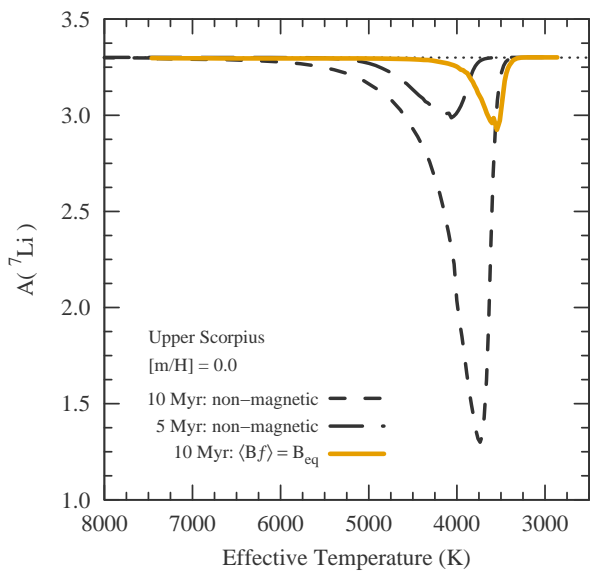}
    \caption{Lithium abundance  as a function of effective temperature at 10 Myr. Predictions are shown from standard (light-blue dashed line) and magnetic (yellow solid line) stellar evolution isochrones. Models were computed using the setup described in Sect.~\ref{sec:models}. Lithium abundance is defined as $A(\,^7{\rm Li}) = \log_{10}(N_{\rm Li}/N_{\rm H}) + 12$.}
    \label{fig:lithium}
\end{figure}

An approach to validating magnetic stellar models is the determination of lithium abundances \citep{MacDonald2015}, or observation of lithium equivalent widths. Magnetic inhibition of convection produces a cooler temperature structure in fully convective pre-main-sequence stars, thereby delaying lithium burning and preserving lithium to older ages \citep{Ventura1998, DAntona2000, MM10, FC13, Malo2014, MacDonald2015}. Starspots have a similar affect on lithium burning timescales \citep{Jackson2014b, Somers2015b}, which is perhaps not surprising given the deep connection between starspots and magnetic inhibition of convection \citep{Biermann1941, Deinzer1965}. Nevertheless, if magnetic fields are directly influencing the structure of young stars, they should leave an imprint on the observed lithium abundance distribution in young stellar populations.
 
Figure \ref{fig:lithium} demonstrates the impact that magnetic inhibition of convection has on predicted lithium abundances for stars in Upper Scorpius at 10 Myr. The lithium depletion boundary at cooler temperatures remains largely unchanged. However, lithium depletion is not expected to occur in warmer stars with effective temperatures exceeding about 4500~K, assuming they possess a strong magnetic field. Therefore, K-stars are not expected to exhibit signatures of lithium depletion, in contrast to standard model predictions, which suggest stars with spectral types later than G are expected to show signatures of lithium depletion. 

Stars with effective temperatures around 3550~K are predicted to exhibit the greatest amount of lithium depletion, as compared to a prediction of 3750~K for standard models. Some depletion is also expected to occur for stars with $3300 <$ \teff/K $< 3500$, which standard models suggest should not be observed. This prediction shows a strong similarity to trends in lithium equivalent widths for M-stars in Upper Scorpius \citep{Rizzuto2015}. The observed lithium equivalent width minimum is located between \teff\ = 3300 -- 3400~K \citep{Rizzuto2015}. While comparing lithium equivalent widths and predictions of lithium abundances is not completely correct, within the narrow range of temperatures surrounding the lithium abundance minimum, equivalent widths should accurately reflect trends in overall lithium abundances. Accounting for uncertainties in the SpT-\teff\ conversion, observed lithium equivalent widths support the validity of magnetic models and an inferred age of 10 Myr for the low-mass stars in Upper Scorpius.

\section{Summary \& Conclusions}
\label{sec:tellit}
Stellar evolution models and literature data are used to estimate median ages for stars with spectral types A through M in Upper Scorpius. A median HR diagram age of 9 -- 10 Myr is found for A-, F-, and G-type stars, with the exception of a population of F-type stars that appear roughly 4 -- 5 Myr older. These results agree with the revised $11\pm3$ Myr age proposed by \citet{Pecaut2012}. At the same time, an HR diagram age of 4 -- 5 Myr is confirmed for K- and M-type stars using standard stellar evolution isochrones \citep{Preibisch2012, Slesnick2008, Herczeg2015}. Notably, isochronal ages of K- and M-type stars in eclipsing multiple-star systems observed by \emph{Kepler}/K2 \citep{Kraus2015, Alonso2015, David2016} disagree with the HR diagram analysis and suggest ages between 6 and 8 Myr for the same stars.

Using stellar evolution models that account for magnetic inhibition of convection \citep{FC12b, FC13}, an age of 9 -- 10 Myr is found for K- and M-type stars in Upper Scorpius. Magnetic inhibition of convection consistently explains:
\begin{enumerate} 
	\item the age discrepancy between high- and low-mass stars in the HR diagram \citep{Herczeg2015},
	\item the observed slope of the low-mass stellar mass-radius relationship,
	\item stellar age differences observed between HR diagram and mass-radius relationship determinations \citep{Kraus2015},
	\item and the lithium depletion pattern inferred from observed lithium equivalent widths \citep{Rizzuto2015}. 
\end{enumerate}
Critically, average surface magnetic field strengths are determined \emph{a priori} via equipartition arguments, highlighting the potential predictive power of magnetic stellar models for young stars.

Results from this work and \citet{Pecaut2012} strongly support the adoption of an approximately 10 Myr median age for the Upper Scorpius subgroup of the Scorpius-Centaurus OB association. There is no population of stars in the Upper Scorpius that unambiguously supports an age of 5 Myr. Nevertheless, additional eclipsing binary systems are needed to better populate the mass-radius relationship to solidify Upper Scorpius as a 10 Myr old stellar association and to verify predictions from magnetic stellar evolution models across the full mass spectrum.

Further work is needed to explore whether magnetic inhibition of convection serves as a robust solution for noted age gradients as a function of stellar spectral type \citep[e.g.,][]{Naylor2009, Herczeg2015}. Previous work determining ages for the $\beta$-Pictoris moving group from magnetic stellar models suggests the solution is robust \citep{MM10, Malo2014, Binks2016}, but additional confirmation using stellar populations with a wider variety of ages is required. These results also highlight the need to explore how the inclusion of magnetic inhibition of convection affects the (sub)stellar initial mass function for young stellar populations.

In the meantime, the use of magnetic stellar models as a component for isochrone analyses for young low-mass stars is recommended. Consistency across the HR and mass-radius diagrams exhibited by magnetic stellar evolution isochrones suggest that magnetic isochrones will provide a more accurate characterization of transiting exoplanet host stars in young stellar populations, especially those observed by \emph{Kepler}/K2 in Upper Scorpius. By adopting an incorrect age, a reasonable estimate of stellar radii can be found, but stellar masses will be systematically underestimated by roughly 30\%. Similar arguments apply to the characterization of young low-mass stars hosting directly imaged giant planets or brown dwarfs. Ages inferred from standard model isochrones may be systematically too young by a factor of two resulting in a systematic underestimation of giant planet and brown dwarfs masses by roughly 35\% \citep{Baraffe2003}. To facilitate their adoption, all models used in this study are available online.\footnote{\url{https://github.com/gfeiden/MagneticUpperSco/}} A small grid of solar-metallicity magnetic stellar mass tracks and isochrones is also under construction, as is a web-interface to compute custom magnetic models.

\begin{acknowledgements}
G.A.F. thanks Alexis Lavail, Eric Mamajek, James Silvester, and Eric Stempels for reading and commenting on an early version of this manuscript as well as Bengt Gustafsson and Thomas Nordlander for interesting discussions that helped improved the manuscript. G.A.F also thanks the anonymous reviewer for posing interesting questions that led to significant improvements in the manuscript. The magnetic Dartmouth stellar evolution code was originally developed with support of National Science Foundation (NSF) grant AST-0908345. This work made use of NASA's Astrophysics Data System (ADS), the VizieR catalogue access tool by CDS in Strasbourg, France \citep{Ochsenbein2000}, and IPython with Jupyter notebooks \citep{Perez2007}. Figures in this manuscript were produced with Gnuplot~5 \citep{Gnuplot5.0}.
\end{acknowledgements}

\end{document}